\documentclass[]{emulateapj}

\slugcomment{Submitted to ApJ}
\shorttitle{The Dark Disk of the Milky Way}
\shortauthors{Purcell et al.}

\begin{document}

\title{The Dark Disk of the Milky Way}

\author{
Chris W. Purcell\altaffilmark{1},
James S. Bullock\altaffilmark{1}, and
Manoj Kaplinghat\altaffilmark{1}
}

\altaffiltext{1}{
Center for Cosmology, Department of Physics and Astronomy, The University of California, Irvine, CA 92697 USA
}

\email{cpurcell@uci.edu}

\begin{abstract}  
Massive satellite accretions onto early galactic disks can
lead to the deposition of  dark matter in disk-like configurations
that co-rotate with the galaxy.  This phenomenon has
potentially  dramatic consequences for dark matter detection experiments.
 We utilize focused, high-resolution simulations of accretion events onto disks
designed to be Galaxy analogues, and compare the resultant disks to the morphological and
kinematic properties of the Milky Way's thick disk in order to bracket the range of
co-rotating accreted dark matter.  We find that 
 the Milky Way's merger history must have been unusually quiescent
compared to median $\Lambda$CDM expectations
and therefore its dark disk must be relatively small:   the fraction 
of accreted dark disk material near the Sun 
is about $20 \%$ of the host halo density or smaller 
and  the co-rotating dark matter fraction
near the Sun, defined as particles moving with a rotational velocity 
lag less than 50 km/s, is enhanced by 
about $30 \%$ or less 
compared to a standard halo model.  
Such a dark disk could contribute dominantly to the low energy (of order
keV for a dark matter particle with mass 100 GeV) nuclear recoil event rate of
direct dectection experiments, but it will not change the likelihood
of detection significantly. These dark disks provide testable
predictions of weakly-interacting massive particle dark matter models
and should be considered in detailed comparisons to experimental
data. Our findings suggest that the dark disk of the  Milky Way may
provide a detectable signal for indirect detection experiments,
contributing up to about $25\%$ of the dark matter self-annihilation
signal in the direction of the center of the Galaxy, lending the
signal a noticeably oblate morphology.   

\end{abstract}

\keywords{Cosmology: theory --- galaxies: formation --- galaxies: evolution}

\maketitle

\section{Introduction} 

Prospects for the direct detection of dark matter depend crucially on
the phase space distribution of dark matter near  the Sun
\citep{Smith_Lewin90,Jungman_etal96,Vogelsberger_etal09}. Unfortunately,
detailed predictions for the dark matter distribution around the
Earth are extremely difficult to construct from first principles; they
require an understanding of spatial clumping on the scale of the
Solar System \citep{Kam08,Peter09}, and are almost certainly
affected by poorly-understood baryonic processes like the formation of
the Galactic disk.   

Recently, \citet{Read_etal08,Read_etal09} emphasized that the process
of cosmological disk galaxy formation can significantly alter the dark
matter distribution compared to canonical predictions that rely on
simulations of the standard halo model.  During the process of
hierarchical structure formation, merging satellite galaxies can get
dragged into the plane of  their host disk and deposit their dark
matter in a structure, dubbed the {\em dark disk} \citep{Read_etal08},
that is co-rotating with the Milky Way stellar disk and
morphologically resembles a thick disk.  If the $\Lambda$CDM cosmology
represents the correct model of structure formation in the universe,
it is certain that dark disks are virtually ubiquitous in disk galaxies.
However, this contribution relative to the smoother halo component will
depend sensitively on the formation process of each galaxy
individually.  In this work, we provide the first focused attempt at 
constraining the dark disk contribution in the Galaxy, improving 
upon the initial discussion of the Milky Way dark disk in \citet{Read_etal08} 
by comparing our simulations to the detailed kinematic and morphological 
properties of the Galactic thick disk.

If a substantial dark disk exists within the Milky Way, then it has
dramatic implications for the direct and indirect detection of dark matter, 
should that species consist of Weakly
Interacting Massive Particles (WIMPs), which are pervasive
in models of new physics at the weak scale. A co-rotating disk of dark
matter will leave temporal modulation signals in terrestrial 
nuclear-recoil experiments.  It will also increase WIMP capture in the
Sun and Earth and enhance the resultant flux of neutrinos from WIMP
self-annihilation; both of these results have been discussed in detail by
\citet{Bruch_etal08,Bruch_etal09}. We show in this work that a disk-like structure could be discernible
in the self-annihilation signal (depending on the total flux) from the 
center of the galaxy. Since the scale height of the dark matter disk
will be larger than the thin disk scale height and smaller than any background
halo flattening, such a morphological
feature could provide an important handle on this indirect detection signal. 

The fraction of dark matter locked up in a co-rotating component is
expected to depend sensitively on merger history  \citep{Read_etal09},
and this creates an important link between dark matter detection
experiments and efforts in Galactic astronomy to constrain the
accretion history of the Milky Way.  Evidence is mounting that the merger
history of the Galaxy is unusually quiescent compared to 
{\em typical} $\Lambda$CDM expectations \citep{Wyse09}.
High-resolution simulations have shown that 1:10 mass-ratio accretion events,
despite being fairly common in the $\Lambda$CDM paradigm
\citep{Stewart_etal08}, are irreconcilable with the cold and thin
Galactic disk \citep[][hereafter PKB09]{Purcell_etal09}.   
This conclusion is consistent with the results of
\citet{Read_etal08}  and \citet{VH08}, who used simulations to show that
1:10 mergers generate heated systems that are grossly consistent galactic {\em thick disks},
but that are clearly incommensurable with the dominant thin disk of the Galaxy.
More recently, the simulations of \citet{Moster_etal09}
showed that while these common mergers
may be consistent with the
broad population of galaxies, they produce remnants that are
clearly thicker than
the Milky Way within $\sim 500$ pc of the disk plane.
Empirically, the Galaxy appears to be deficient in stellar mass and
angular momentum when balanced against a sample of local spiral
galaxies \citep{Hammer_etal07}, further motivating the need
for a focused program that constrains the Milky Way's dark disk
specifically, rather than relying on general cosmological expectations.  

\begin{figure}[!t]
\centerline{\epsfxsize=3.4in \epsffile{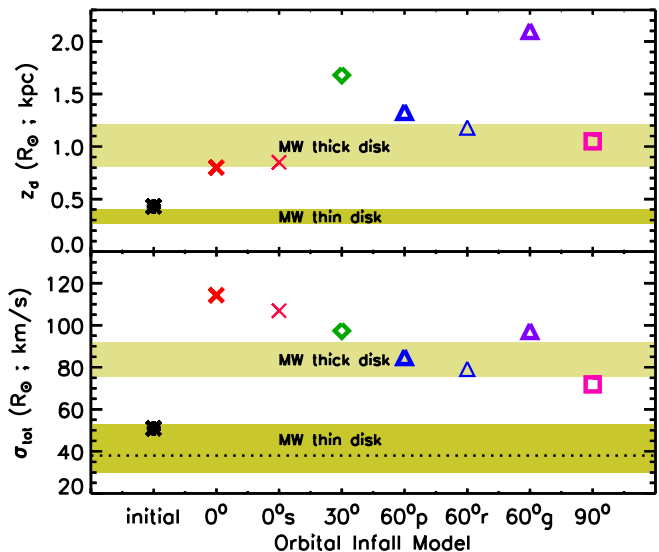}}  
\caption{Morphological and dynamical properties of the final stellar disks at the solar neighborhood for each fiducial 
orbital infall condition: the disk scale height $z_d$ in kpc ({\em upper} panel), and the total velocity dispersion 
$\sigma_{\rm tot}$ in km/s ({\em lower} panel).  The shaded regions indicate observational boundaries: the Galactic 
scale heights and their errors are drawn from the SDSS tomographic results reported by \citet{Juric_etal08}, while the 
dynamical temperature ranges represent the thin-disk spread observed by \citet{Nordstrom_etal04} for total velocity 
dispersion as a function of stellar age in the solar neighborhood (the region represents stars with ages between 1 Gyr 
and 8 Gyr and the {\em dotted} line marks the velocity dispersion for stars with a median age of $\sim2-3$~Gyr), and 
the thick-disk range corresponds to the dispersions reported by \citet{Soubiran_etal03}.  The infall model denoted by 
$0^{\circ}{\rm s}$ represents the accretion event involving a slower satellite galaxy; the three $60^{\circ}$ infall 
models are marked with prograde, retrograde, and gas, according to the methods outlined in \S\ref{sec:methods} and as 
listed in Table~\ref{accretedtable}.}
\label{fig:properties}
\end{figure}

We extend the methodology of PKB09 and use high-resolution simulations to study dark disk production.
Our main observational constraint comes from observed kinematic and spatial properties of the old 
($\sim 10$ Gyr) thick disk of the Milky Way, which comprises at most $\sim 20 \%$ of the total mass of 
the Galaxy \citep[{\em e.g.}][]{Juric_etal08}.  We initialize stellar disks that are consistent with the observed 
properties of the Milky Way today and consider the impact of fairly massive, cosmologically common 
accretion events.  This approach is conservative: by initializing a disk that is as massive as the Milky Way 
disk today, we are exploring a case that is both more resistant to heating and more efficient in capturing 
dark matter from the accreted satellite than any realistic progenitor of the Milky Way's thick disk.  More 
generally, by focusing on the oldest, thickest, and hottest stellar component of the Galactic disk, we are 
able to provide a most conservative constraint on the Milky Way's merger history.  

The heating of the initial disk and the creation of the dark disk are intimately connected, since heating 
requires the transfer of kinetic energy from the incoming satellite to the disk.  For a fixed set of orbital 
parameters, more massive accretion events produce more stellar heating, and they also deposit more 
dark matter into a disk-like configuration.  The transfer of kinetic energy depends on how close the 
satellite gets to the disk and its orbit about the disk; a closer and slower orbit results in larger heating 
and concomitantly larger stripping of dark matter from the satellite by the tides of the Milky Way.  Thus, 
by comparing the heated stellar disk to the observed Milky Way thick stellar disk, we are able to bound 
the amount of coherently rotating dark matter that is deposited during the disruption of a large satellite 
galaxy.

As alluded to above, the Milky Way has both a thin and thick disk \citep{GR83}, 
with the latter component being significantly hotter and thicker than the former.  
By comparing the stellar remnants in our simulations with those of the Galactic 
thick disk implicitly, we allow for the possibility that the thin disk of
the Milky Way is regrown later through fresh accretion.  While 
the regrowth of a thin disk could potentially act to make the heated disk 
thinner, it cannot make the heated disk colder.  Thus our constraints based 
on the resultant velocity ellipsoid are the most robust. 

The majority of the accretion events we consider have mass ratios
$M_{\rm sat}$:$M_{\rm host} = 1$:$10$.  This type of merger represents
the dominant mode by which dark matter halos grow in
$\Lambda$CDM models \citep{Purcell_etal07,Stewart_etal08}. They are 
also the most relevant for the formation of co-rotating dark disks, because 
more massive mergers are both quite rare and catastrophically
destructive to primary disks, while smaller mergers are less heavily affected 
by dynamical friction and leave behind much less dark matter (we explore 
such a case below).  Thus the 1:10 events we consider here are both
cosmologically common and capable of producing a co-rotating dark disk 
without completely destroying the primary galaxy.  
Note that while a number of less massive mergers may incrementally build
up a thick disk \citep[see][]{Read_etal08}, they will not produce a coherently 
rotating dark disk unless the angular momenta contributed by the individual 
accretion events also coincide constructively.  This tendency toward a series of
prograde accretions is not found in $\Lambda$CDM halos, at least for
the few merger histories that have been explored in detail \citep[][]{Kazantzidis_etal08}.


\begin{figure*}[!t]
\centerline{\epsfxsize=6.5in \epsffile{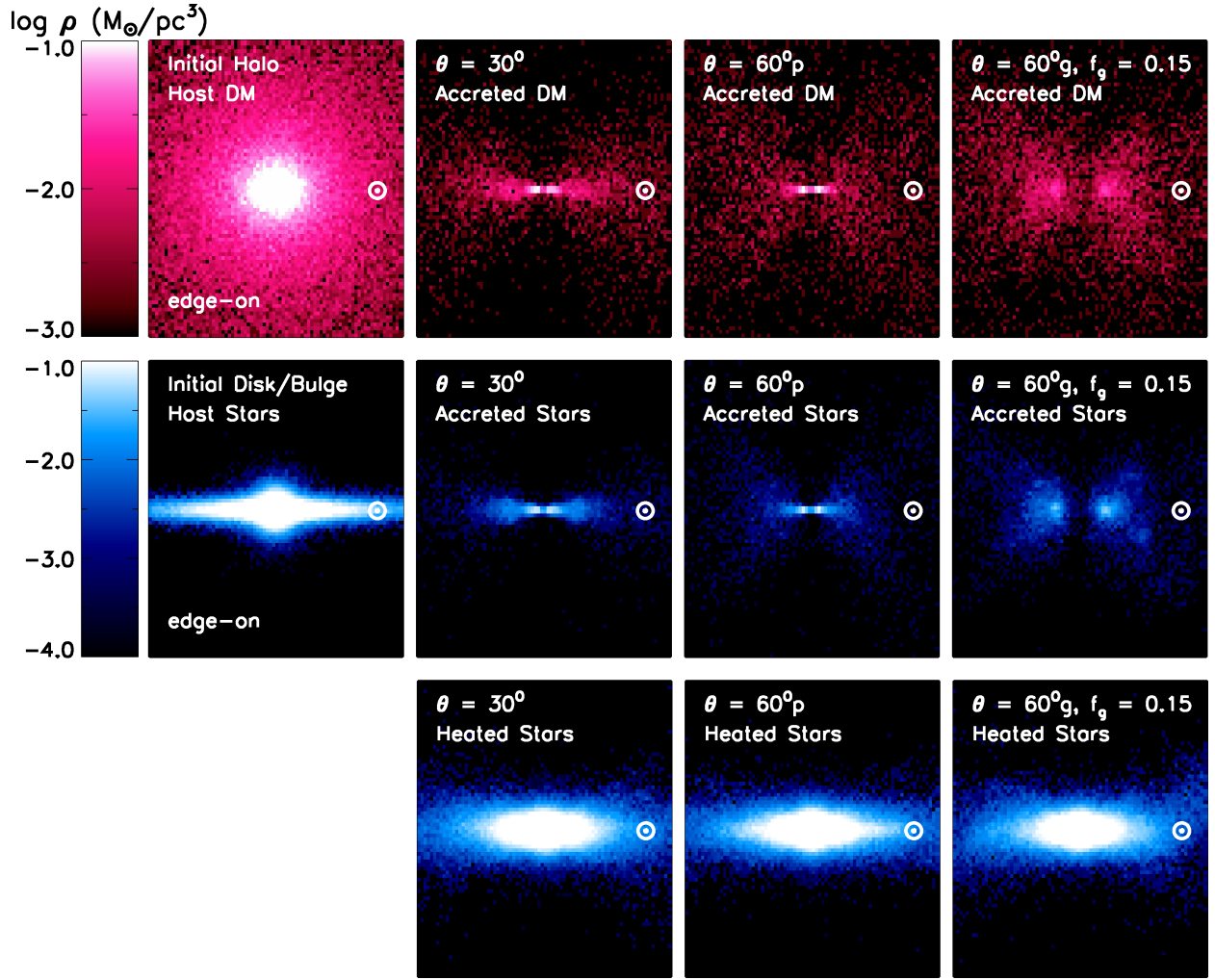}}
\caption{The three-dimensional dark matter density ({\em top} panels) and stellar density ({\em middle} panels) 
contributed by the accreted satellite galaxy for a selection of orbital conditions, viewed edge-on with respect to the 
primary galactic stellar disk and calculated in a vertical slice with a thickness of 1 kpc which passes through the 
center of the galaxy.  For reference, the host galaxy's structure is mapped in the {\em left} panel of each figure group 
and the density maps of the heated primary disk stars are shown in the {\em lower} panels; 
each box is 20 kpc on a side and the solar position is marked with $\odot$.  
}
\label{fig:densitymaps}
\end{figure*}

In addition to collisionless experiments, we perform a hydrodynamical
simulation in which the primary galaxy also hosts a gaseous disk, 
\citep[a much more thorough treatment of these and similar results 
can be found in Kazantzidis, Purcell, \& Bullock 2009, in preparation; see also][]{Moster_etal09} 
to determine whether this component plays a substantial role in the
formation of a dark disk.  We find that our results do not depend sensitively 
on the presence of gas as we have modeled it, though we caution that the 
treatment of ISM gas physics in galaxy simulations remains uncertain.   Based 
on our implementation, the presence of gas does not curtail the heating of a 
primary disk that occurs during dark disk creation.

\section{Methods}
\label{sec:methods}

In a suite of high-resolution collisionless simulations, PKB09 investigated the response of a galactic disk to the 
infall of cosmologically-common accretion events involving satellites with one-tenth the mass of the host halo.  For our primary system, 
we focus here on the Galaxy 1 model from PKB09, which is a good match to the Milky Way today, and should provide a 
higher probability of dark disk creation via satellite dragging than any presumably less-massive progenitor of the 
Galaxy at higher redshift.  We now provide a brief overview of our simulations and refer the reader to PKB09 for a more 
complete discussion of the methods and results regarding the morphological and dynamical changes undergone by stellar 
disks in response to these accretion events.


\begin{figure*}[t]
\centerline{\epsfxsize=6in \epsffile{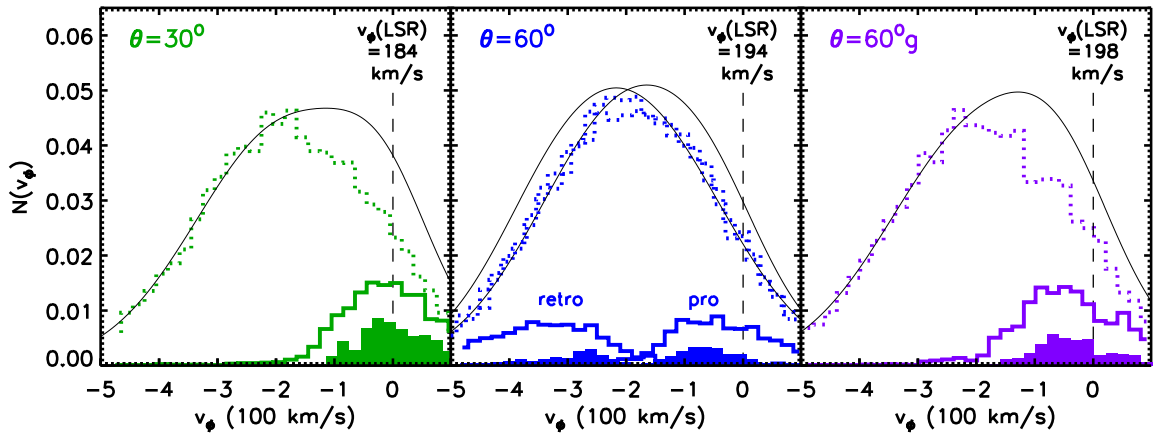}}  
\caption{Rotational velocity histograms at the solar radius for accreted dark matter ({\em thick solid}), accreted stars 
({\em shaded region}), and host halo dark matter ({\em thin dotted}) for infall inclinations corresponding to those shown 
in Figure~\ref{fig:densitymaps}, calculated with respect to the local standard of rest defined by the rotation speed 
$v_{LSR}$ of the primary galactic stellar disk.  The dark matter values are normalized in each bin to the total number of 
dark particles in the area defining the solar neighborhood; the number of accreted stars in each bin is normalized to the 
total number of stars in the same region.  In each panel, the {\em thin solid} line represents the sum of the two 
Gaussians representing dark matter in the host halo and from the accreted subhalo.}
\label{fig:velocity}
\end{figure*}

Our primary galaxy is constructed according to the fully self-consistent distribution functions prescribed by 
\citet{Widrow_etal08} and is therefore an equilibrium solution to the coupled collisionless Boltzmann and Poisson 
equations.  The model's stellar disk is initialized with an exponential scale length $R_d=2.84$~kpc and a vertical 
distribution described by a sech$^2$ function with scale height $z_d=0.43$~kpc, containing $10^6$ particles with a 
total disk mass $M_{\rm disk} = 3.6 \times 10^{10} M_{\odot}$.  The massive central bulge follows a S\'ersic profile 
of effective radius $R_e=0.58$~kpc and index $n=1.118$, and has a mass $M_{\rm bulge}=9.5 \times 10^9 M_{\odot}$ 
distributed among $5 \times 10^5$ particles.  These stellar components are embedded in a dark host halo composed of 
$4 \times 10^6$ particles which follow the canonical NFW density profile of \citet{Navarro_etal96}, with scale radius 
$r_s=14.4$~kpc and virial mass $M_{\rm host} \simeq 10^{12} M_{\odot}$.  This particular set of parameters was chosen 
in order to minimize secular effects such as bar formation, as well as artificial heating induced by the interaction of 
disk particles with more massive halo particles.  

The abscissa of Figure~\ref{fig:properties} as well as the top row of Table~\ref{accretedtable} denote the infall 
models we consider. Every simulation tracks the infall of a satellite galaxy 
modeled within a subhalo of virial mass $M_{\rm sat} \simeq 10^{11} M_{\odot}$ at $z=0.5$.  The satellites are each composed 
of $9 \times 10^5$ particles in a density structure well fitted by an NFW profile with concentration 
$c_{\rm vir} \simeq 14$, and contain a stellar mass $M_{\star} = 2.2 \times 10^9 M_{\odot}$ distributed among $10^5$ 
particles in a spheroid with S\'ersic index $n \sim 0.5$.  The stellar masses for the satellite galaxies are drawn from the model of 
\citet{Conroy_Wechsler09}, which describes the redshift-dependent connection between the observed spatial abundance of 
galaxies and the corresponding abundance of predicted dark matter halos.  Fiducially, we investigate five orbital infall 
inclinations, including four prograde orbits with $\theta=(0^{\circ},30^{\circ},60^{\circ}$p, and~$90^{\circ})$ and 
one retrograde $60^\circ$r orbit.  The satellites are initialized 120 kpc from the host halo center, and 
we set our initial orbital vectors according to the distribution of substructure 
accretions drawn from cosmological simulations \citep{Benson_05,Khochfar_Burkert06}, with radial and tangential velocity 
components equal to $v_r=116$~km s$^{-1}$ and $v_t=77$~km s$^{-1}$ respectively.  Additionally, in a model we designate 
$0^\circ$s, we investigate a case where the same satellite galaxy is traveling more slowly than in the fiducial 
$0^\circ$ case, with $v_r=58$~km s$^{-1}$ and $v_t=38.5$~km s$s^{-1}$; these orbital parameters correspond roughly to the lower 
$1\sigma$ limit of \citet{Benson_05}. 

To test the effect of a full treatment of hydrodynamics on the morphological and dynamical changes induced by an 
accretion event, we repeat the simulation involving a massive satellite galaxy infalling along a prograde orbit with an 
inclination of $60^{\circ}$ (hence the designation $60^{\circ}$g), having converted a modest fraction $f_g=15\%$ 
of the initial primary galaxy's star particles into a gaseous component with the same density structure as the stellar disk.  
Our hydrodynamical prescription includes atomic cooling for a primordial mixture of hydrogen and helium, and the star 
formation algorithm is based on the work of \citet{Katz92}, in which gas particles in cold and dense regions form star 
particles at a rate proportional to the local dynamical time; star formation occurs when the gas density exceeds 0.1~cm$^{-3}$ 
and the gas temperature drops below $1.5\times10^4$~K.  Supernova feedback is implemented according to the blast-wave 
model described in \citet{Stinson_etal06}, in which the energy deposited by a Type-II supernova into the surrounding gas is 
$4\times10^{50}$~ergs. This parameter set produces realistic galaxies in cosmological simulations \citep{Governato_etal07}.  
Again we refer the reader to Kazantzidis, Purcell \& Bullock (2009, in preparation) for more details regarding the suite of 
experiments to which this current test case belongs.

As a test of the fractional contribution made by a much smaller satellite galaxy, we simulate an additional accretion 
event with identical orbital parameters to the fiducial infall, involving a subhalo of mass 
$M_{\rm sat} \sim 4 \times 10^{10} M_{\odot}$, i.e. with mass ratio 1:25.  For this case, we choose a planar subhalo 
orbit ($\theta = 0^{\circ}$) in order to maximize the likelihood of dark disk formation.  However, this experiment 
resulted in a trivial increase ($\lesssim 1\%$) in the fraction of dark matter in the solar neighborhood, and is thus not listed 
in Table~\ref{accretedtable}.

 \section{Results}
\label{sec:results}

\subsection{Gross Comparisons to the Galactic Stellar Disk}

We summarize in Figure~\ref{fig:properties} the observable properties of our initial 
and remnant stellar disks, with each model indicated by name along the horizontal 
axis.  The upper panel shows the disk scale height $z_d$, derived from fitting the minor axis surface density profile 
at a defined solar radius ($R_\odot = 8$ kpc), using the two-component form 
\begin{equation}
\Sigma(z) = \Sigma_{\rm d} \, {\rm sech}^{2}(z/z_{\rm d}) + \Sigma_{\rm diffuse} \, {\rm sech}^{2}(z/z_{\rm diffuse}).
\end{equation}
This decomposition provides a minimum estimate of the resultant disk scale height ($z_d$) by allowing for a secondary, much 
thicker component ($z_{\rm diffuse}$), which dominates at large height and low surface brightness.  Generally, we find significant 
variance in these values with orbital inclination angle; the disk scale height $z_d \sim 1-2$~kpc, while the faint component 
typically has $z_{\rm diffuse} \sim 4-7$~kpc.  Note that the initial thin disk of the primary galaxy has been destroyed: our 
minimal scale heights are comparable to or larger than the Galactic thick disk scale height, as demonstrated in the comparison of 
Figure~\ref{fig:properties}, where observed scale heights for the Milky Way thin and thick disks are shown as shaded horizontal 
bands.  These values are drawn from \citet{Juric_etal08}, in which an exponential disk scale height is presented; we convert this value here to a sech$^2$ scale height 
such that the two vertical surface density profiles decrease by roughly the same amount within $\sim 1$ scale height of the 
disk plane\footnote{Note that much of the relevant literature, including {\em e.g.} \citet{Moster_etal09}, uses the typical conversion factor 
of two, which is only valid at heights much larger than the scale height of the disk.  Our conversion method captures more accurately 
the behavior of the surface brightness profile at heights comparable to and smaller than the disk scale height.}.  
The total three-dimensional velocity dispersion $\sigma_{\rm tot} = (\sigma_R^2 + \sigma_\theta^2 + \sigma_\phi^2)^{1/2}$ of each disk at the solar 
neighborhood is presented in the bottom panel of Figure~\ref{fig:properties}.  The significant increase in this quantity from the 
initial model to the resultant disk betrays the extraordinary heating undergone by the primary disk during the merger.  The shaded 
bands again reflect observed velocity dispersions for the Milky Way thin and thick disks, as drawn from \citet{Nordstrom_etal04} 
and \citet{Soubiran_etal03}, respectively.


\begin{figure}[t]
\centerline{\epsfxsize=3.7in \epsffile{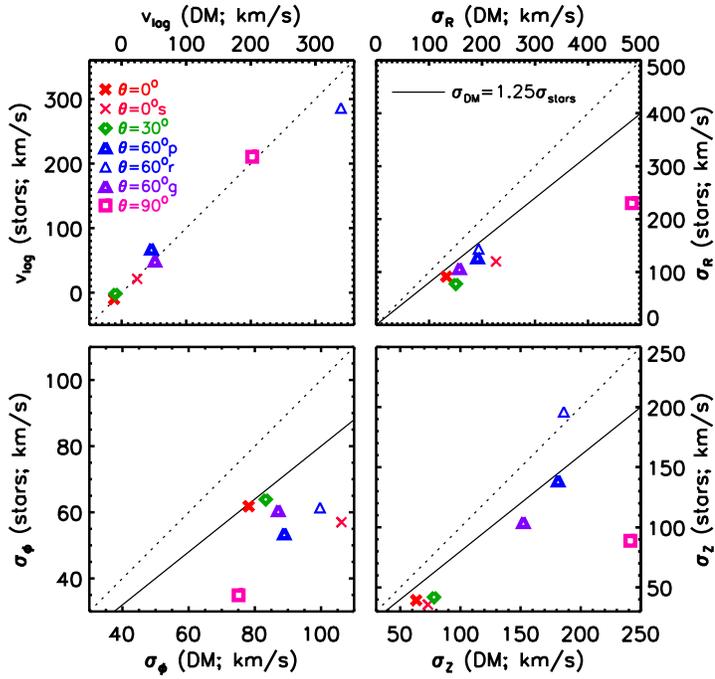}}  
\caption{Correlations between the kinematic properties of dark matter and stars accreted onto the primary galaxy during 
satellite infall.  The {\em upper left} panel shows the amount $v_{\rm lag}$ by which each species of accreted material 
lags the rotational speed of the primary galactic disk; negative (positive) quantities indicate faster (slower) rotation 
than the local standard of rest as defined by the mean stellar velocity of disk stars in the solar neighborhood.  The 
remaining three panels plot each axis of the velocity ellipsoid for both species.  In all figures, the ordinate marks 
the values for accreted stars, while the abscissa denotes the corresponding values for accreted dark matter; the 
{\em dotted} line defines where the two values are equal, while the {\em solid} line forms the upper bound 
$\sigma^{DM} = 1.25 \sigma^{stars}$ of an envelope which roughly contains the simulation results. }
\label{fig:correlation}
\end{figure}


\begin{figure}[!t]
\centerline{\epsfxsize=3.4in \epsffile{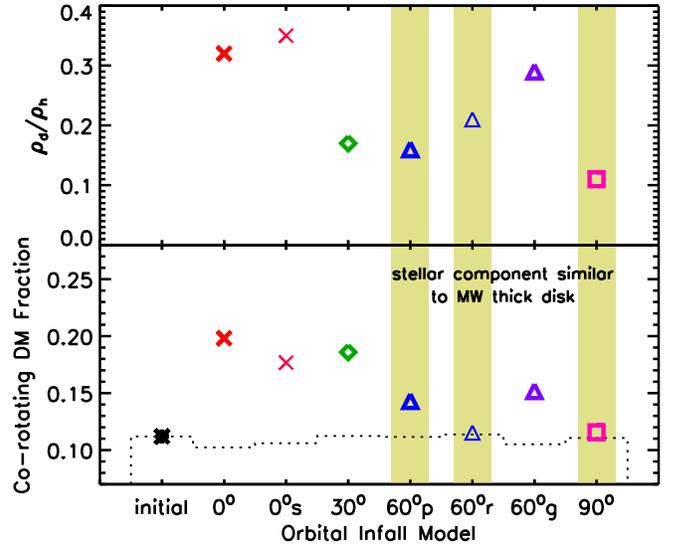}}  
\caption{Similar to Figure~\ref{fig:properties}, morphological and dynamical properties of the final accreted dark disks 
at the solar neighborhood for each fiducial orbital infall condition: the local density ratio $\rho_d/\rho_h$ ({\em upper} 
panel), and the fraction of local dark matter moving slowly ($|v_{\rm lag}| \leq 50$~km/s) with respect to the stellar 
disk's rotational speed ({\em lower} panel).  In the latter case, the {\em thin dotted} histogram denotes the contribution 
to the fraction of slow-moving dark matter made by the initial host halo, while the colored points indicate the sum of 
this component and the amount of co-rotating dark matter deposited by the accreted satellite galaxy.  The shaded regions 
correspond to the orbital infall models that produce stellar remnants with morphology and kinematics similar to the 
Galactic thick disk, as demonstrated by Figure~\ref{fig:properties}. }
\label{fig:darkproperties}
\end{figure}

If we associate our remnant disks with the Milky Way thick disk, most of our resultant systems are 
thicker and hotter than the Galaxy.  This comparison is particularly constraining for two reasons.  First, the age 
distribution of stars in the {\em thin disk} includes stars that are 8-10 Gyr old, suggesting than any merger of this 
kind must have happened well before $z \sim 1$ (when the primary disk was even smaller and less able to cause significant 
dragging into the disk plane).  Moreover, while regrowth of a new disk could potentially reduce the scale height of the 
thickened disk, it would not reduce the velocity dispersion of the stars.  Only the prograde and retrograde $60^{\circ}$ 
orbits and the $90^\circ$ degree orbits result in a system that is marginally viable, both morphologically and dynamically.

\subsection{Morphological and Kinematic Characterizations of the Dark Disk}

Each of the merger simulations provides an accreted dark component and an accreted stellar component, but the nature of 
these components depends sensitively on the interaction, as illustrated in Figure~\ref{fig:densitymaps} for three of our 
simulations.  For reference, the upper left panel presents a central slice of the dark matter density in the primary 
galaxy's host halo prior to the accretion, the middle left panel shows a slice in the stellar distribution in the 
primary galaxy, and the lower panels represent the heated stars belonging to the original disk only.  The three columns 
on the right for the upper two rows display the resultant distributions of accreted dark matter and accreted stars; 
as might be expected, the lower-latitude event produces the most disk-like accreted dark matter morphology, with a 
qualitatively similar accreted stellar distribution that is somewhat more rotationally supported owing to the fact that 
the stars are more tightly bound in the satellites than the dark matter.  As noted in Table~\ref{accretedtable}, the 
low-latitude simulations also produce the largest fractional density $\rho_d/\rho_h$ of accreted dark matter compared to 
background halo dark matter in the solar neighborhood\footnote{Throughout this work, we define the solar neighborhood as 
the area between galactocentric radii $7<R<8$~kpc and bounded vertically by $|z| \leq 2$~kpc; we also note that variance 
in these boundaries does not lead to significantly altered results.}.  High-latitude events create less discernible dark 
disks at the solar position and deposit proportionally less dark matter there; we also note that our test model involving 
the planar infall of a satellite galaxy with mass ratio $M_{\rm sat}$:$M_{\rm host} = 1$:$25$ evolves over a much longer 
timescale than that of the fiducial models, and also results in a negligible contribution to the local dark matter 
quotient, indicating that even a multitude of small subhalos with similar orbits may not be able to form a significant 
dark disk.  For the models with mass ratio $M_{\rm sat}$:$M_{\rm host} = 1$:$10$, Table~\ref{accretedtable} provides dark 
disk and accreted stellar disk scale heights at the solar neighborhood, as computed by fitting a sech$^2$ profile.
   
An additional characterization of interest is the velocity distribution of the dark disk component.  It is useful to 
parametrize the dark matter in the solar neighborhood of our simulations with a double Gaussian distribution that 
represents the original host and accreted subhalo material:
\begin{equation}
f (v_R, v_\phi, v_z) =  f_{\rm host}(v_R, v_\phi, v_z) + f_{\rm acc}(v_R, v_\phi, v_z),
\end{equation}
where in each term
\begin{equation}
f(v_R, v_\phi, v_z) = A  \, e^{-\sum(v_i-\bar v_i)^2/2\sigma_i^2}.
\end{equation}
The sum involves each velocity component, $i=R,\phi,z$.  We work in a frame that co-rotates with the resultant stellar 
disk such that the host halo produces a net dark matter wind in the $\phi$ direction that is close to the primary galaxy 
rotation speed $(\bar{v}_\phi)_{\rm host}  \sim -200$ km/s and $\bar{v_R} = \bar{v_z} \sim 0$.  Accreted matter can 
co-rotate or anti-rotate with the stellar disk and we quantify this rotation in accreted material by the lag speed 
$v_{\rm lag} =  - \bar{v}_\phi$, such that a zero lag corresponds to precise co-rotation, while a positive lag trails 
the stellar disk and a negative lag means that the dark disk is rotating faster than the stellar disk.  The resultant 
velocity distribution parameters for each accreted component in our simulations are presented in Table~\ref{accretedtable}. 

In Figure~\ref{fig:velocity}, we show the collapsed distributions in rotational velocity $v_\phi$ for the background host 
halo as well as accreted dark matter, along with our double-Gaussian fit to the total amount of dark matter in 
the solar neighborhood, for the same three resultant galaxies that were illustrated in Figure~\ref{fig:densitymaps}: 
$30^{\circ}$, $60^{\circ}$p, and $60^{\circ}$g.  The center panel also presents the velocity distribution for the 
$60^{\circ}$r retrograde merger.  The filled histograms in each panel show the velocity distributions for accreted stars, 
normalized relative to the total stellar mass at the solar location.  Two points are immediately apparent; first, that 
lower-latitude accretion events produce smaller lag velocities in both accreted species, and secondly that the lag speed 
of accreted stars grossly mimics that of the accreted dark matter for accretion events at all latitudes.

\subsection{Correlating the Dark and Stellar Disks}

The kinematic relationship between the accreted dark and stellar disks \citep[a phenomenon 
also reported by][]{Read_etal08,Read_etal09} is important because it may provide an avenue to constrain the dark disk velocity 
distribution via observational studies that can isolate an accreted stellar component.  In Figure~\ref{fig:correlation}, 
we show the correlations between accreted dark matter and accreted stars in each velocity dispersion component, as well as 
the rotational lag speed, which correlates very well between the two accreted species.  We also note that the velocity 
dispersion of accreted stars generally provides a lower limit on the velocity dispersion of the dark disk in each 
directional component; the correlations in Figure~\ref{fig:correlation} indicate that dark matter dispersions are consistently 
at least a factor of $1.25$ larger than their stellar counterparts.  

Though not explicitly shown here, we also find a connection between the velocity-ellipsoid axis 
ratio $\sigma_z/\sigma_R$ of the accreted dark matter and that of the stars formerly belonging to the satellite galaxy, in 
that the two collisionless events involving the high inclination angle of $\theta=60^{\circ}$ both have vertical to radial 
velocity dispersion ratios of approximately unity for all accreted material, while the velocity ellipsoids resulting from the low-latitude 
events are significantly more oblate, with both stellar and dark axis ratios roughly equal to $\sim 0.5$: a similar value 
to that obtained observationally for the Galactic solar neighborhood \citep{Gomez_etal90,Nordstrom_etal04}.  It should be 
noted here that the polar infall results in a final velocity-ellipsoid axis ratio coincident with that of the low-latitude 
events, due to a very large dispersion in the radial velocity of both accreted species in this case.  

\subsection{Summary of Results}

We use two simple metrics to characterize our main results in Figure~\ref{fig:darkproperties}; as in 
Figure~\ref{fig:properties}, each model is indicated along the horizontal axis.  The upper panel presents the accreted 
dark matter fraction $\rho_d/\rho_h$ in the solar neighborhood; the lower panel presents the fraction of local dark 
matter that is roughly co-rotating with the stellar disk, which is potentially the most salient characteristic of an 
accreted dark disk.  Specifically, we define the latter quantity to be the fraction of accreted dark particles with 
rotational speeds within 50 km/s of the stellar disk's velocity, {\em i.e.} $|v_\phi - v_{LSR}| \leq 50$~km/s.  Note 
that by this definition, even the initial spherical halo has a non-zero co-rotating dark matter fraction; for reference, 
the dotted line in the lower panel of Figure~\ref{fig:darkproperties} shows the fraction of dark matter particles initially 
from the background halo that are co-rotating with the stellar disk in the final simulation snapshot.

As might be expected, the most significant dark disks form during the planar accretion events $0^\circ$ and $0^\circ$s,
which respectively produce $32\%$  and $35\%$ accreted dark matter fractions in the solar vicinity and factors of $1.7$ and 
$1.4$ enhancements in the co-rotating dark matter fraction.  However, these planar events also produce stellar disks that 
are much hotter than the thick disk of the Milky Way (as shown in Figure~\ref{fig:properties}), indicating that these cases 
are not appropriate baseline models for the Galaxy.  Though not evident in Figure \ref{fig:properties}, the slower orbit actually 
produces more radial heating than the fast orbit (see Table~\ref{accretedtable}) and is therefore more discrepant with the 
Milky Way thick disk velocity structure than the standard orbit.

The three shaded bands in Figure~\ref{fig:darkproperties} highlight the three model cases that produce stellar disks 
marginally consistent with the thick disk of the Milky Way, when considering morphology as well as dynamical temperature.  
The most optimistically viable model for a dark disk is the 60-degree prograde orbit, which results in a $16 \%$ accreted 
dark matter contribution and a $14 \%$ co-rotating dark matter fraction; the latter quantity represents a $\sim 30 \%$ increase 
compared to the co-rotating dark matter fraction in the smooth halo case.  As noted in Table~\ref{accretedtable}, the dark 
disk scale height in this case is $z_d = 4.6$~kpc and lags the LSR with $v_{\rm lag} = 46$ km/s. 

\begin{figure*}[!t]
\centerline{\epsfxsize=6.5in \epsffile{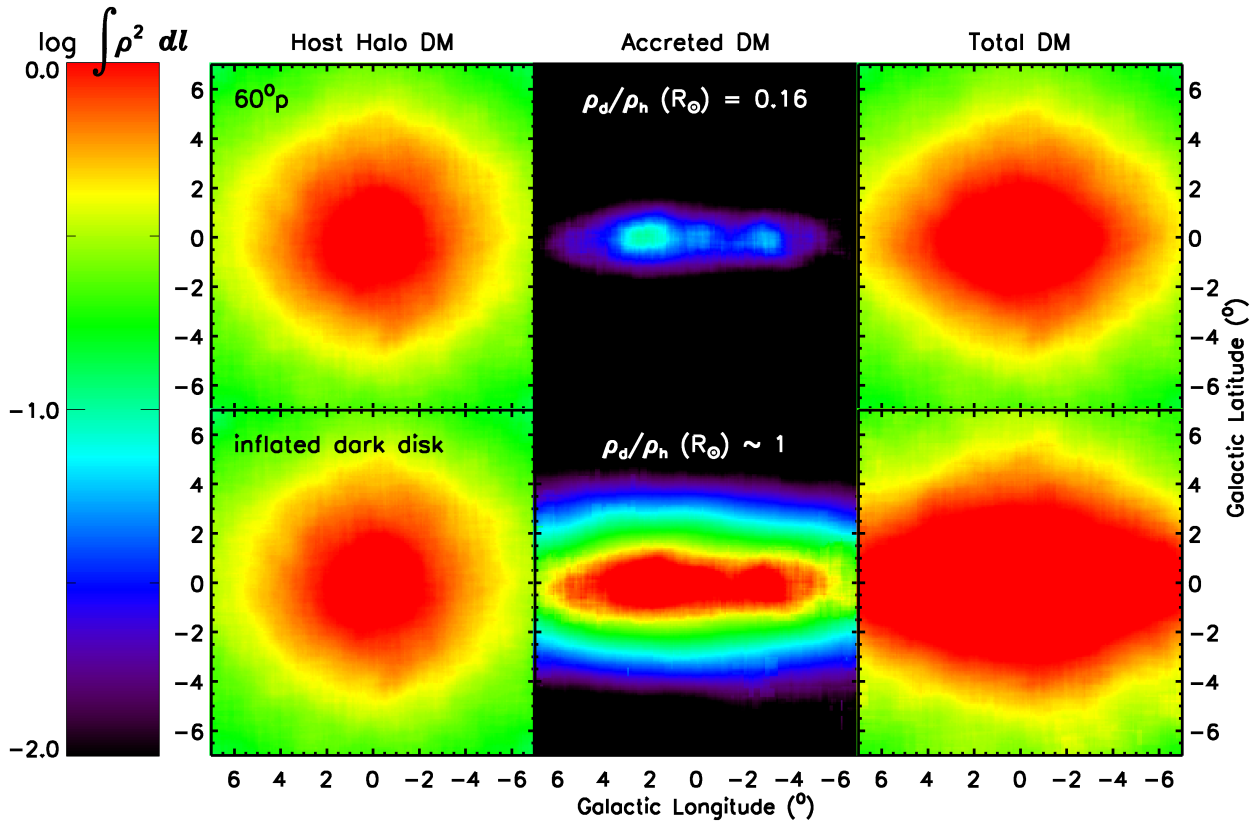}}
\caption{The dark matter annihilation signal in Galactic coordinates,
  mapped with a color logarithmically proportional to the integrated
  square of the particle density along a line of sight toward the
  center of the galaxy, following the prograde satellite galaxy infall with
  inclination $\theta=60^{\circ}$.  Note that the combined signal 
  ({\em right} panels) is noticeably more oblate than the host material 
  alone ({\em left} panels; shown {\em after} the accretion event); 
  the accreted dark matter ({\em center} panels) contributes roughly $\sim 10-25 \%$ 
  of the annihilation rate at Galactic longitudes $\sim 5^{\circ}-10^{\circ}$.  In 
  the lower set of figures, we test the effect on the total annihilation 
  signal for a hypothetical dark disk with a local density 
  ratio at the solar neighborhood $\rho_d/\rho_h \sim 1$; this boost is 
  also an $O(1)$ effect, contributing $\sim 50\%$ of the total signal over the 
  same range in longitude.  We have argued that such a large dark 
  disk component would likely have to form through secular processes 
  unrelated to satellite accretion events. 
}
\label{fig:annihilation}
\end{figure*}

Although the thick disk we produce is obviously more massive than the Galactic thick disk, the 
similarity in morphology and kinematics motivate us to move beyond limits, and 
speculate on the properties of the actual dark disk in the Milky Way.  
Given the range of parameters probed here and in \citet{Read_etal08}, and operating under the assumption that the 
Milky Way's thick disk was formed in an event similar to the type we simulate here, we expect that the real Galactic 
dark disk contributes roughly $10\%$ to the local dark matter density.  As we discuss below in \S\ref{sec:summary}, such a 
dark disk does not significantly enhance the possibility of direct detection of dark matter; however, it will be important 
to include in any detailed analysis aimed at interpreting an observed WIMP detection signal.  Moreover, such a dark disk 
may provide a detectable morphological signature towards the Galactic center in experiments designed to indirectly detect 
dark matter via its self-annihilation products.

 \section{Comparison to Previous Work}
\label{sec:compare}

In a recent effort to quantify disk heating in a $\Lambda$CDM context, \citet{Read_etal08} used simulations similar to those 
presented here in order to show that massive satellite accretion events involving a Galactic-type thin disk not only produce thick 
stellar systems but also result in the deposition of accreted dark matter into a disk-like component that co-rotates to some degree 
with the primary galaxy.  Specifically, they analyzed dark disk formation during the infall of various subhalo types (of particular 
interest here are their LMC and LLMC models, representing satellite galaxies with total mass $M_{\rm sat} \simeq 2.4 \times 10^{10}$ 
and $1.0 \times 10^{11}$ respectively), finding that for low-inclination mergers, the resulting density ratio of accreted dark matter 
to that of the background host halo in the solar neighborhood was $\rho_d/\rho_h = 0.22$ for the smaller subhalo and $0.42$ 
for the more massive satellite that corresponds closely to our fiducial infalling system.  In the follow-up analysis of cosmological 
simulations of Milky-Way-type disk galaxies \citep{Read_etal09}, the authors found similar results for systems with a wide range in 
accretion history activity, showing that several massive mergers at late times can result in a dark disk with local density similar to that 
of the host halo itself.

Although our simulated phenomena are qualitatively similar to those presented by \citet{Read_etal08}, there are two 
key points of distinction to be drawn.  The two methods are only significantly discrepant in the choice of initial radii and velocities 
of the infalling satellites.  Our work adopts satellite galaxy velocity vectors drawn from subhalo infall distributions found in cosmological 
simulations, and thus we set the initial position of the subhalo far ($\sim 120$~kpc) from the host halo's center in order to minimize 
the sudden change in potential felt by the primary galactic disk and also to imitate the cosmological velocity conditions, which are 
measured at the host's virial radius.  In contrast, the infalling satellites of \citet{Read_etal08} begin much closer ($\sim 30$~kpc) to the 
primary disk's center and involve subhalos traveling at speeds slower than our fiducial initial velocity by roughly fifty percent; 
slightly faster than our $0^{\circ}$s case.  The authors motivate the choice of a slow orbit at a small initial Galactocentric apocenter by 
suggesting it replicates the orbital parameters of satellites found in a loose group environment that is accreted onto a Galaxy-sized host.  
However, it still remains to be shown that such a velocity vector at $\sim 30$~kpc can arise naturally for a massive satellite within a 
cosmological setting.  Moreover, any process that would act to bring a subhalo to such a low energy state in the disk plane would almost 
certainly heat the primary disk to a degree unallowable by observations.

The difference in initial conditions may be largely responsible for the small systematic differences between our simulated 
dark disk results and those of \citet{Read_etal08}.  While our maximal local density ratio is $\rho_d/\rho_h \sim 0.35$, 
they find $0.42$ in their experiment labeled LLMC-$10^{\circ}$, an accretion event involving a subhalo of similar mass.  
The methodological variance may also explain why our disks are slightly dynamically hotter.  Our 
$0^\circ$ case has $\sigma_{\rm tot} \sim 117$~km/s and our $0^\circ$s case has $\sigma_{\rm tot} \sim 110$~km/s,
while their LLMC-$10^{\circ}$ model has $\sigma_{\rm tot} \sim 105$~km/s.  It is important to emphasize, however, that both sets 
of simulations find resultant thick disks that are hotter than the observed thick disk of the Milky Way, which has 
a total dispersion of just $\sim 85$~km/s. Therefore, in a broad sense, we agree that these disks are not ideal analogues to the Milky Way.

We note that a more substantial dark disk component could emerge following multiple similar accretion events, as investigated in cosmological 
simulations of Galaxy-sized disks by \citet{Read_etal09}, but such a series of late-time mergers cannot be reconciled with the relatively thin disk 
of the Milky Way.  Unfortunately for experiments aimed at detecting local dark matter, the quiescent Galactic accretion history favored by our 
result indicate that the relevant observable quantities are probably far less affected by a dark disk than we might have hoped.

 \section{Discussion and Interpretation}
\label{sec:summary}

In our experiments, we have adopted assumptions throughout that maximize 
the likelihood for producing a dark disk remnant and for preserving a thin, cold disk by initializing a primary disk that is as massive as the Milky Way disk 
today. This approach is quite conservative because we compare the heated stellar remnant to the old {\em thick disk} of the
Milky Way, which is at least five times less massive than our primary system \citep[{\em e.g.}][and references therein]{Juric_etal08}.
We have argued that the dark disk formed in our $60^\circ$ prograde merger case provides a fair limit on the properties of an 
underlying accreted dark component of our Galaxy, with a local contribution $\rho_d/\rho_h \simeq 0.15$ that is at the low end of dark disk 
contributions estimated by \citet{Read_etal08} for Milky Way type galaxies in $\Lambda$CDM.  

Based on this limit, we expect that nuclear recoil experiments
designed to directly detect local dark matter would only receive small
boosts to the event rate, preferentially at lower energies.  
Dark disk fractions of $\rho_d/\rho_h \sim 10 \%$ were explored by
\citet{Bruch_etal08}, who found an order unity enhancement in the
differential event rate at keV recoil energies for a 100 GeV dark
matter particle. For a TeV mass dark matter particle, such an
enhancement would be present at higher energies (accessible to
CDMS-II).  While an order unity enhancement will not significantly impact the
event rate, \citet{Bruch_etal08} showed that the phase of the annual
modulation signal is sensitive to both the dark disk and the mass of
the dark matter particle. The phase of the annual modulation signal
produced by the dark disk depends on the motion of the Sun relative to
the dark disk. The phase of the modulation of the total event rate
in a given energy window depends on both the mass of dark matter
particle and the dark disk fraction. If the dark disk fraction is
bracketed from the lower end, this will result in a prediction for the
phase given a dark matter particle mass. This may be testable at 
future detectors if a positive signal is seen.  In addition, these
results also imply that directional detectors should see the WIMP 
wind direction change as a function of recoil energy, the details of
which would depend on the energy window, the mass of the WIMP, the
dark disk fraction and the motion of the Sun with respect
to the dark disk. 

An increase in the local number density of particles that co-rotate
with the Sun will also enhance WIMP capture in the Sun and the Earth.
These captured particles self-annihilate into standard model particles,
including neutrino pairs. Thus a larger co-rotating fraction of dark
matter will result in a larger neutrino flux potentially observable with
experiments like Super-Kamiokande and IceCube.   
In detail, the capture rate depends sensitively on $\rho_d/\rho_h$,
the dark disk lag speed $v_{\rm lag}$,  and the velocity dispersion of
the dark disk.  The timescale for
capture of WIMPs at Earth is small compared to the age of the planet
and hence the WIMP density in the Earth has not yet reached
equilibrium. In the Sun, as shown by \citet{Bruch_etal09}, the
situation is the opposite for most regions of parameter space and
the number of WIMPs would have reached an equilibrium value. Thus the
annihilation rate of WIMPs inside the Earth is proportional to the
capture rate squared, while in the Sun it is equal to half the capture
rate (given the equilibrium condition there).  \citet{Bruch_etal09} show that this
leads to two to three orders of magnitude enhancement in the flux from
the Earth and an order of magnitude enhancement in the flux from the
Sun for $\rho_d/\rho_h=1$ and an assumed isotropic Gaussian velocity
distribution for the dark disk with lag $v_{\rm lag}=50$~km/s and 1-D
dispersion $\sigma_d=v_{\rm lag}$. They also consider dark disk  
parameters $\rho_d/\rho_h=0.25$ and $\sigma_d=100$ km/s and state that 
this does not lead to a large boost in the signal, owing primarily to
the large velocity dispersion. By comparison, our simulated prograde 
accretion event with $\theta=60^{\circ}$ results in a local density
ratio of $0.15$, velocity dispersion $\sigma_\phi \simeq 90$~km/s,
and similar lag speed $v_{\rm   lag} \simeq 50$~km/s. Thus we do not
expect large boosts to the neutrino flux from the Earth or the Sun,
and our results imply that the likelihood of WIMP detection through
this indirect channel is not significantly increased  due to an
accreted dark disk. 

Although the prospects for direct detection and indirect detection via
WIMP capture are not significantly altered by the accreted dark disk, 
we note  that the accreted dark matter distribution is
relatively more prominent toward the center of the galaxy, where  
most of the satellite galaxy's material settles.  The center of the
Galaxy is a prime target for experiments aimed at detecting dark
matter indirectly via high energy annihilation  products
\citep{Bergstrom_etal98}, as the flux scales proportionally to
the square of the WIMP density.  In Figure~\ref{fig:annihilation}, we
view the halo's center from the vantage point of the solar
neighborhood, after the prograde $60^{\circ}$ satellite infall, and
estimate the annihilation signal at Earth by integrating $\rho^2$
along lines of sight.  We notice immediately  that while the host's
dark matter has retained approximate sphericity (left),  the combined
signal (right) which includes the accreted subhalo material (middle)
has  a significantly more oblate contour.  Approximately $\sim
10-25\%$ of the  total $\rho^2$ contribution at galactic longitudes
between $5^{\circ}$  and $10^{\circ}$ comes from accreted dark
matter.  As in the case of direct detection, the dark disk does not
significantly change the likelihood of detection; however, if there is a
positive detection, then our results provide motivation to search for
a disk-like component with scale height and length that are
different from that of the Milky Way thin disk. 

Our focus in this work has been on the scenario of \citet{Read_etal08}, who 
showed that dark disks could be created via satellite accretion events.  We have 
argued that under this scenario, the relative importance of the accreted 
dark disk can be constrained via detailed comparison to the properties of thick 
disk stars in the Galaxy and that this constraint limits the dark disk's density 
contribution to $\sim 0.2 \rho_{\rm halo}$ in the solar neighborhood.  
However, it is interesting to consider the theoretical possibility that a dark 
disk with density ratio $O(1)$ could form in response to some other process 
that did not heat the disk.  For example, one could imagine that large infalling 
gas clouds could transfer angular momentum to the dark matter, creating a 
disk-like dark component without disturbing the primary disk significantly.  
Such a process may not be easy to arrange, given the large amount of angular 
momentum transfer required to produce a co-rotating dark disk as well as the 
fact that observed disk galaxies seem to require something close to angular 
momentum conservation.

Nonetheless, it is an instructive exercise to determine the effect of
an $O(1)$ dark disk on indirect  detection signals.  To investigate
this, we have scaled the local density ratio for  the $60^{\circ}$
prograde case such that $\rho_d/\rho_h =1$, and the resultant
annihilation signal's sky map is shown in the lower panels of
Figure~\ref{fig:annihilation}.   The disk-like nature of the signal is
clearly apparent and could be easily distinguished if any dark matter
self-annihilation is observed toward the center of the Galaxy.
Therefore, if such a dark disk could somehow form without involving
satellite accretion events, it would be discernible through indirect
detection experiments.  Coupled with the boost in the direct
detection signal that is expected for such an $O(1)$ dark disk
\citep{Bruch_etal09}, the jointly observable consequences of such a
component would be significant.  

In a similar vein, it is worth noting that the quantity of interest
for direct detection experiments is the fraction of slow-moving dark
matter particles. From Figure~\ref{fig:darkproperties}, we see that
the $30^\circ$ case shows an order unity enhancement in slow-moving
dark matter particles, and Figure~\ref{fig:properties} shows that this
run produces a disk that is only marginally hotter than the thick disk
of the Milky Way. However, the scale height of the disk is
considerably larger. One may, however, hypothesize that a later stage
of disk formation could change the disk scale height substantially
without heating the stars further. Therefore, we urge the reader to
keep in mind that $O(1)$ enhancement in slow-moving particles is
possible in the accreted dark disk scenario of \citet{Read_etal08},
but that the expected velocity dispersion in the $\phi$ direction is
still large -- in our $30^\circ$ case, it is 83 km/s (see
Table~\ref{accretedtable}).   

In this work we have argued that within the context of the accreted dark disk scenario 
of \citet{Read_etal08}, it is likely that the dark disk of the Milky Way contributes 
approximately $10-20\%$ to the local dark matter density.  If so, then its presence 
may be important to include in any attempt to interpret a detection signal of WIMP dark 
matter in the solar neighborhood and for the indirect detection signal from the Galactic 
center.  Given this, it will be important to constrain the dark disk to a higher degree of 
accuracy.  Both \citet{Read_etal08} and we have demonstrated  
that there is a significant dynamical relationship between dark matter
accreted during the infall of a massive subhalo and the stellar mass
contributed by that satellite galaxy to the primary system
(Table~\ref{accretedtable} and  Figure~\ref{fig:correlation}).
Therefore one can anticipate constraining the properties of any Milky
Way dark disk by isolating a dominant subpopulation of accreted
Galactic disk stars, perhaps by some combination of chemical and
dynamical tags.  For example, if we were to consider the unlikely
scenario where the entire thick disk of the Milky Way was deposited
by a single accretion event ({\em i.e.} it is devoid of any
pre-existing disk stars), then we may consider the dynamical
properties of the Galaxy's thick disk as a means to constrain the dark disk:
$(\sigma_R,\sigma_\phi,\sigma_z) =  (63\pm6,39\pm4,39\pm4)$~km/s,
$v_{\rm lag} = 51.5$~km/s \citep{Soubiran_etal03}, and  $z_d =
0.90\pm0.18$~kpc \citep{Juric_etal08}.  Under this interpretation, our
simulation results would imply that the lag speed of the MW dark disk
is $\sim 50$ km/s and that it must be quite hot, with 1.25
$(\sigma_R,\sigma_\phi,\sigma_z) \sim  (88, 51, 62)$ km/s, and with a
very thick $z_d \gtrsim 2$ kpc.  These numbers are almost certainly
lower limits; as we have shown in Table~\ref{accretedtable}, accreted
stellar distributions tend to be both hotter and thicker than the
primary disks they heat.  If there is any pre-existing disk
population associated with the thick disk of the Milky Way, the
accreted portion is almost certainly hotter than the composite thick
disk values used in our analysis. 

Although theoretical predictions based on galaxy formation models as
they relate to dark matter detection are still in the nascent stage,
it is clear that the standard halo model is insufficiently equipped to
allow precise predictions for measurable quantities of interest;
scattering rates and flux scalings depend crucially on the presence of
locally coherent substructural flows.  Observations have thus far
been unable to accurately estimate the degree to which the Galactic
thick  disk's population is composed of stars stripped from a
satellite galaxy during tidal disruption, as opposed to material
heated to ejection from the thin stellar disk \citep[as investigated
by][]{Purcell_etal09b}.  In principle, however, a combination of
high-precision kinematic measurements and detailed chemical
composition data should be able to define a  phase-space in which
accreted and original stars occupy disparate regions.  Placing such
dynamical constraints on the  population of stars obtained by the
Milky Way during a significant accretion event may enable us to
identify the kinematics of the dark matter deposited during that same
event, thereby placing more stringent limits on the availability of a
dark Galactic disk for the purpose of detection experiments.   

\begin{table*}[t] 
\centering
\begin{minipage}{\textwidth}
\centering
\tabcolsep 6pt
\caption{Properties of accreted dark matter and stars in the solar neighborhood}
\label{accretedtable}
\begin{tabular}{@{}ccccccccc}
\hline
\hline
Quantity & & & & Orbital Inclination & & &  & \\
\hline
(DM) &  $\theta = 0^{\circ}$ & $\theta = 0^{\circ}$-slow & $\theta = 30^{\circ}$ & $\theta = 60^{\circ}$  &  $\theta = 60^{\circ}$-gas  &  $\theta = 60^{\circ}$-retro  &   $\theta = 90^{\circ}$    & Units\\
\hline        
$\rho_d/\rho_h$  & $0.32$   &   $0.35$     & $0.17$      & $0.16$    & $0.29$          & $0.21$       &  $0.11$    & \\ 
$\sigma_R$        & $132.7$     & $226.5$    & $150.5$    & $191.6$   & $157.4$    &  $193.5$   &  $483.0$           & [km/s]   \\
$\sigma_{\phi}$    & $78.2$ & $106.2$  & $83.3$   & $88.9$   & $87.1$    &  $99.7$       &  $75.1$                & [km/s]\\
$\sigma_z$          & $63.4$  & $73.1$ & $78.1$   & $181.2$   & $152.3$    &  $185.9$   &   $241.2$              & [km/s]  \\
$v_{\rm lag}\footnote{Lag velocities are calculated in the frame of the local standard of rest (LSR), defined by the mean rotational speed of the primary galaxy's disk stars at the solar location; positive (negative) values indicate a slower (faster) rotation than that of the LSR.}$     & $-11.7$ &  $24.1$  & $-9.6$ & $46.1$    & $51.5$   & $339.6$       &  $202.2$      & [km/s] \\
$z_d$     & $2.6$  & $1.8$   & $4.2$           & $10.9$             & $9.2$      &  $9.2$     &  $36.6$   & [kpc]     \\
\hline 
(stars)\footnote{For reference, the Milky Way's thick disk is described by $(\sigma_R,\sigma_\phi,\sigma_z) =  (63\pm6,39\pm4,39\pm4)$~km/s,
$v_{\rm lag} = 51.5$~km/s \citep{Soubiran_etal03}, and  $z_d =
0.90\pm0.18$~kpc \citep{Juric_etal08}.}  
 & & & & & & &  & \\
\hline        
$\sigma_R$   & $91.5$   & $119.9$    & $77.6$    & $127.6$    & $107.2$       &  $143.7$   &  $230.1$        & [km/s]   \\
$\sigma_{\phi}$   & $61.8$   & $57.0$    & $63.9$   & $53.6$  & $60.6$        &  $61.4$       &  $34.9$         & [km/s]\\
$\sigma_z$      & $39.3$  & $35.7$  & $41.9$      & $138.9$          & $104.3$        &  $196.1$   &   $88.8$         & [km/s]  \\
$v_{\rm lag}$     & $-9.8$    &   $21.3$   & $-1.6$   & $68.3$    & $49.5$          & $286.2$       &  $210.6$    & [km/s] \\
$z_d$     & $1.7$  & $0.8$   & $1.6$           & $4.6$             & $2.8$      &  $9.4$     &  $11.5$  & [kpc]     \\
\hline
\hline
\end{tabular}
\end{minipage}
\end{table*}

\acknowledgments

We thank Justin Read,  Laura Baudis, Tobias Bruch, and George Lake for discussions that 
improved the accuracy and clarity of our presentation.  
Stelios Kazantzidis assisted in the construction of
of the initial conditions used in this work, and we thank him for allowing us to present the results 
of our collaborative disk-satellite experiments here, including the hydrodynamical experiment 
described above.  We also thank Larry Widrow and John Dubinski for kindly making available 
the software used to set up the initial galaxy models. 
CWP and JSB are supported by National Science Foundation (NSF) 
grants AST-0607377 and AST-0507816, and the Center for 
Cosmology at UC Irvine.  MK is supported by NSF grant AST-0607746 and 
NASA grant NNX09AD09G.  The numerical simulations 
were completed primarily on the IA-64 cluster at the San Diego Supercomputing Center, 
with ancillary experiments performed on the GreenPlanet cluster at UC Irvine.

\end{document}